\documentclass[apj]{emulateapj} 

\newcommand{\millimeter}{millimeter} 
\newcommand{\dpt}{DPT00~2}
\newcommand{\irsione}{IRS-I~1}
\newcommand{\irsonee}{IRS1E}
\newcommand{\irsitwo}{IRS-I~2}

\newcommand{\methanol}{CH$_3$OH}
\newcommand{\ammonia}{NH$_3$}
\newcommand{\ntwohplus}{N$_2$H$^+$}
\newcommand{\lsun}{$L_\odot$}
\newcommand{\msun}{$M_\odot$}
\newcommand{\ngc}{NGC~6334}
\newcommand{\ngcf}{NGC~6334~F}
\newcommand{\ngciabbrev}{NGC~6334~I}
\newcommand{\ngci}{NGC~6334~I}
\newcommand{\ngcin}{NGC~6334~I(N)}
\newcommand{\mjb}{mJy~beam$^{-1}$}
\newcommand{\HII}{H\,{\sc ii}}

\begin{document}

\shortauthors{Hunter et al.}

\shorttitle{Multiplicity in \ngci\ and I(N)}

\submitted{}
\received{02 March 2006}
\revised{14 May 2006}
\accepted{17 May 2006}

\title{Millimeter Multiplicity in NGC~6334~I and I(N)}
\author{T.R. Hunter\altaffilmark{1}, C.L. Brogan\altaffilmark{2},
S.T. Megeath\altaffilmark{1,3}, K.M. Menten\altaffilmark{4}, 
H. Beuther\altaffilmark{5}, S. Thorwirth\altaffilmark{4}
}
 
\email{thunter@cfa.harvard.edu}

\altaffiltext{1}{Harvard-Smithsonian Center for Astrophysics, Cambridge, MA 02138}  
\altaffiltext{2}{NRAO, 520 Edgemont Rd, Charlottesville, VA 22903} 
\altaffiltext{3}{Ritter Observatory, MS 113, University of Toledo, Toledo, OH 43606-3390} 
\altaffiltext{4}{MPIfR Bonn, Germany} 
\altaffiltext{5}{MPIA Heidelberg, Germany} 

\begin{abstract}

Using the Submillimeter Array (SMA), we have imaged the 1.3~millimeter
continuum emission at the center of the massive star-forming regions
\ngci\ and I(N).  In both regions, the SMA observations resolve the
emission into multiple millimeter sources, with most of the sources
clustered into areas only 10,000~AU in diameter.  Toward \ngci, we find
four compact sources: the two brightest (I-SMA1 and I-SMA2) are
associated with previously-known ammonia cores; I-SMA3 is coincident
with the peak of the compact \HII\/ region (\ngcf); and I-SMA4 is a
newly-discovered object.  While I-SMA3 exhibits a mixture of free-free
and dust emission, the rest of the objects are dust cores.  Toward
\ngcin, seven compact dust cores are found, one of which is associated
with a faint centimeter source.  With the exception of I-SMA3, none of
the millimeter sources have infrared counterparts in {\it Spitzer}
Space Telescope 3-8~$\mu$m images.  Using a simple physical model for
the dust continuum emission, the estimated mass of interstellar
material toward each of these compact objects is in the range of 3 to
66 \msun.  The total mass in the compact objects appears to be similar
in I and I(N).  The small size of these groups of sources suggest that
these objects are proto-Trapezia forming in the centers of clusters of
low to intermediate-mass stars.

\end{abstract}


\section{Introduction}

The formation process of massive stars continues to be a
poorly-understood phenomenon in astrophysics.  The most fundamental
clues to the origin of OB stars are their multiplicity and their
common association with high column densities of molecular gas and
dust.  A unique feature of OB stars is that they are often found in
non-hierarchical (and consequently) non-stable systems in the center
of clusters \citep{Sharpless54}.  A nearby example of this phenomenon
is the Orion Trapezium with projected stellar separations of 4000 to
10000 AU, which in turn is found in the center of a rich cluster of
low mass stars.  A long-standing question is whether these multiple
systems are the direct result of the formation process, thus providing
a clue to the process of high mass star formation, or the result of
dynamical evolution in the centers of young clusters where the most
massive stars move toward the center by ejecting lower mass stars
outward \citep{Bonnell98}.  While recent (sub)millimeter studies have
identified good candidates for massive protoclusters on scales of
several parsecs,
the identification of proto-Trapezia is a more difficult prospect.
Due to the complex nature of high mass star-forming regions, the high
extinction typically observed toward massive protostars, and their
typically large distance ($>1$~kpc), high-resolution imaging at
(sub)millimeter wavelengths is required to resolve one protostar from
another.  Recent space-based near-infrared (NIR) imaging has succeeded
in identifying a 5600~AU diameter cluster of five proto-OB stars
comprising W3~IRS~5 \citep{Megeath05}.  However, because massive
protostars form in the deeply-embedded cores of molecular clouds, dust
extinction may obscure a significant fraction of them in the infrared.
In these cases, \millimeter\ continuum emission from dust provides one
of the few alternative tracers of protostars because it remains
optically thin at high column densities ($N_{\rm H} \lesssim 10^{25}
{\rm cm}^{-2}$).

The recent commissioning of the Submillimeter Array (SMA) on Mauna
Kea, Hawaii has expanded the range of \millimeter\ interferometry to
higher frequencies and lower declinations.  Now within reach is \ngc\
($\delta = -35^\circ$), a luminous and relatively nearby
\citep[1.7~kpc;][]{Neckel78} molecular cloud/\HII\/ region complex
containing several concentrations of massive star formation at various
stages of evolution \citep{Straw89a}.  The northeastern end appears to
be the youngest and contains the radio source ``F'' (G351.42+0.64; 
\citet{Rodriguez82})
which is associated with IRAS~17175-3544.  The earliest far-infrared
images of this region \citep{Emerson73,McBreen79} identified the
emission as source ``I''.  An additional component ``I(N)'' was first
detected at 1mm \citep{Cheung78} and later at $400\mu$m
\citep{Gezari82}.  Further observations have demonstrated that
although these two cores have comparable mass, \ngci\ dominates the
combined bolometric luminosity of $2.6\times 10^5$ \lsun\
\citep{Sandell00}.  In the NIR, an embedded cluster of stars has been
detected in the central $2'$ of \ngci\ \citep{Tapia96}.  In contrast,
the only NIR emission detected toward I(N) are H$_2$ knots that are
most likely associated with outflow activity (see Figure~3 of
\citet{Megeath99}).  In this paper, we present the first 1.3
\millimeter\ interferometry of \ngciabbrev\ and I(N).  Despite their
strikingly different appearance in the NIR, we find that both regions
contain a similar cluster of compact dust continuum cores.  A detailed
treatment of the \millimeter\ line emission accompanying these objects
will be the subject of a forthcoming paper (Hunter et al., in
preparation).
 
\section{Observations}

\subsection{Submillimeter Array (SMA)}

The SMA\footnote{The Submillimeter Array (SMA) is a collaborative project
between the Smithsonian Astrophysical Observatory and the Academia
Sinica Institute of Astronomy \& Astrophysics of Taiwan.}
observations were made with six antennas in both the compact
configuration (2004 May), and extended configuration (2005 May).  Two
pointings were observed: \ngci\ at $17^{\rm h}20^{\rm m}53^{\rm
s}.44$, $-35^{\circ}47'02''.2$ and \ngcin\ at $17^{\rm h}20^{\rm
m}54^{\rm s}.63$, $-35^{\circ}45'08''.5$ (J2000).  Unprojected
baseline lengths ranged from 22m to 226m. 
The SMA receivers are double sideband mixers with 2~GHz bandwidth
\citep{Blundell04}.
The center frequencies were 217.6 in LSB and 227.6 GHz in USB. The
correlator was configured for uniform resolution (0.81~MHz per
channel).  Typical system temperatures were 200K. The gain calibrators 
were NRAO~530 ($23\arcdeg$ distant) and J1924$-$292 ($27\arcdeg$ distant) 
and the bandpass calibrators were Uranus and 3C279.
The data were calibrated in Miriad, then exported to AIPS where the line 
and continuum emission were separated with the task UVLSF.
Self-calibration was performed on the continuum data, and solutions
were transferred to the line data.  
Flux calibration is based on SMA flux monitoring of the observed quasar
and is estimated to be accurate to within 20\%.  
The estimated accuracy of the absolute coordinates is $0.4''$.
After combining the calibrated LSB and USB continuum uv-data, the 
1$\sigma$ rms noise level achieved in the continuum images is 7~\mjb\/.
The resulting synthesized beam
is $2\farcs1 \times 1\farcs2$ and the primary beam is $\sim 56\arcsec$.

\subsection{Very Large Array (VLA)}

Archival 3.6 cm data from the NRAO\footnote{The National Radio
Astronomy Observatory is a facility of the National Science Foundation
operated under agreement by the Associated Universities, Inc.} Very
Large Array (VLA) were calibrated and imaged in AIPS.  The observation
date was May 7, 1990 (A-configuration). The flux calibrator was
3C286, and the gain calibrator was NRAO530.  The synthesized beam
of the 3.6 cm continuum image is $0.90''$ $\times$ $0.43''$ at
(P.A.=$-18\arcdeg$) and the rms sensitivity is 1~\mjb\/.

\subsection{{\it Spitzer} Space Telescope}

Both SMA fields were observed with the IRAC camera \citep{Fazio04}
onboard the {\it Spitzer} Space Telescope.  The total integration time
was 666 seconds.  The high dynamic range mode was used to obtain 0.4
and 10.4 second integrations.  The data were reduced using the BCD
data from the Spitzer Science Center version 11.4 pipeline.  When
constructing images, the 0.4 second data were used toward pixels where
the 10.4 second data were saturated, which is only the case for some
of the sources in \ngci.  The pixel size for the images is $1.''2$.
Upper limits were derived from mosaics produced with the MOPEX
program.  To calculate the upper limits, the rms signal was found for
a 5x5 pixel box centered on the SMA sources.  These values were
converted to units of mJy per pixel, then multiplied by a factor of
$5/2$ to calculate $5\sigma$ limits for detections in a 2x2 pixel
aperture.

\section{Results} 

\subsection {\ngci}

\begin{figure*}
\centerline{\includegraphics[scale=0.78]{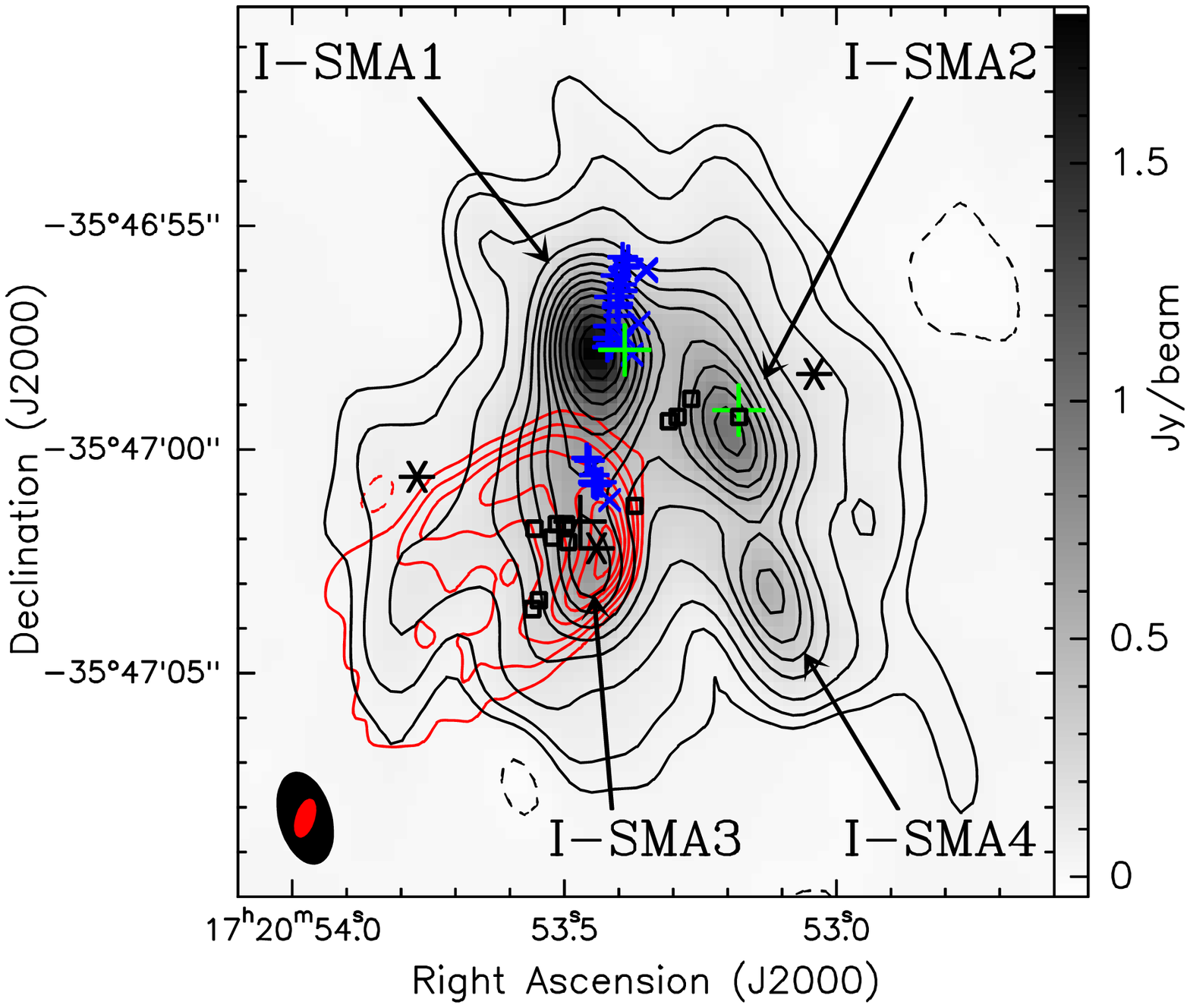}}  
\caption{{\it Grayscale and black contours:} 1.3mm continuum image 
of \ngci\ taken from the combined LSB plus USB data. Contour levels are 
(-3, 3, 6, 12, 18, 24, 36, 48, 64, 80, 100, 120, 150, 180) 
$\times$ 7.5 mJy/beam.  Water masers
are marked by the blue symbols: 
'x' \citep{Migenes99} and '+' \citep{Forster92} (shifted by
the amount suggested in a private communication from J.~Forster 
in \citet{Carral97}).
Open squares mark Class II \methanol\ masers 
\citep{Walsh98}.   
The black cross marks IRS~1E and the six-pointed stars mark the three
mid-infrared sources from east to west: DPT00~2 \citep{DeBuizer00}, 
\irsione\ and \irsitwo\ \citep{DeBuizer02}. The green crosses mark
\ammonia\ peaks (derived from images of the hyperfine lines of 
\citet{Beuther05}). 
{\it Red contours:} VLA 3.6 cm emission 
(-5, 5, 15, 30, 60, 90, 120, 150 \mjb\/).
\label{fig1}}
\end{figure*}

The 1.3 mm continuum emission in \ngci\ is shown in Figure~\ref{fig1}.
We have resolved four major continuum sources which we denote as
I-SMA1..4, in descending order of peak intensity (see
Table~\ref{table1}).  The continuum sources are all fairly compact.
Gaussian fitting yields formal size estimates between 1.6 and 2.7
arcsec (2800 and 4600~AU).  Including compact and extended emission,
the total flux density within Figure~\ref{fig1} is $10.7 \pm 0.1$~Jy,
equivalent to $43 \pm 5$\% of the single dish continuum flux density
\citep{Sandell00}.  The overall extent of the emission ($\approx
10''$) agrees well with the size estimate from single-dish maps
\citep{Sandell00}.  No other sources are found beyond this central
concentration of objects.  The two brightest \millimeter\ sources
(I-SMA1 and I-SMA2) correspond within $0.5''$ to the two primary
\ammonia\ peaks observed in the (1,1), (2,2), and (3,3) transitions
\citep{Beuther05,Kraemer95}.  The H$_2$O masers form a linear
structure \citep{Forster92} that appears to be associated with I-SMA1.
In contrast, the 6.7 GHz \methanol\ masers are associated with I-SMA2
and I-SMA3.  I-SMA3 coincides to within $0.3''$ of the peak of the 3.6
cm emission which traces a cometary compact \HII\/ region
\citep{Carral97}.  Based on the 3.6~cm flux density, we estimate that
the free-free contribution to the 1.3~mm flux density of I-SMA3 is
$\sim 1.25$~Jy (62\%).  The rest of the 1.3~mm continuum emission from
this source (and all of the other sources) originates from dust.
I-SMA3 lies $1.0''$ south of the NIR star \irsonee\ which has been
proposed as the exciting source of the compact \HII\/ region
\citep{Persi96}.  Since free-free emission dominates I-SMA3, it is
perhaps not surprising that its position lies closer to the mid-IR/cm
peak than to \irsonee.  The known mid-IR sources \irsitwo\
\citep{Persi98} and \dpt\ \citep{DeBuizer02} reside within the
extended \millimeter\ contours, but the lack of a compact millimeter
source at these positions supports the hypothesis of
\citet{DeBuizer02} that they are not internally heated.  With the
exception of the \HII\/ region nebula (I-SMA3), none of the
\millimeter\ sources are detected in the IRAC images.  Indeed, no
compact counterpart at any other wavelength has been observed toward
I-SMA4.

\begin{table*} 
\centering
\caption{Properties of millimeter continuum sources in \ngci\ and I(N) \label{table1}}  
\begin{tabular}{cccccccc}  
\hline\hline
Source & \multicolumn{2}{c}{J2000 coordinates} & $I_{\rm 1.3mm}$  & $F_{\rm 1.3mm}$
       & $F_{4.5\mu{\rm m}}$
       & $F_{8.0\mu{\rm m}}$ \\
       &  $\alpha$ ($^{\rm h}~~^{\rm m}~~^{\rm s}$)   & $\delta$ ($^{\circ}~~{'}~~{''}$) & (Jy/b) 
       & (Jy) & (mJy)  & (mJy)
        \\
\hline
I-SMA1 & 17 20 53.44 & -35 46 57.9  & $1.77 $ & $3.49 \pm 0.70$ &  $<320$ & ..\tablenotemark{a}\\  %
I-SMA2 & 17 20 53.20 & -35 46 59.6  & $0.96 $ & $2.28 \pm 0.46$ &  $<240$ & $<1490$ \\  %
I-SMA3 & 17 20 53.45 & -35 47 02.6  & $0.71 $ & $2.01 \pm 0.40$ &  ..\tablenotemark{a} & ..\tablenotemark{a}\\  %
I-SMA4 & 17 20 53.12 & -35 47 03.2  & $0.52 $ & $1.10 \pm 0.22$ &  $<260$ & $<1490$ \\ %
\hline  
Total &              &               &  & $10.7 \pm 2.1$\tablenotemark{b} & \\
\hline

I(N)-SMA1 & 17 20 55.21 & -35 45 04.1 & $0.82$ & $2.04 \pm 0.41$ &  $<0.11$ & $<0.71$ \\ 
I(N)-SMA2 & 17 20 54.90 & -35 45 06.8 & $0.35$ & $0.50 \pm 0.10$ &  $<0.35$ & $<0.53$ \\ 
I(N)-SMA3 & 17 20 55.00 & -35 45 07.5 & $0.27$ & $0.39 \pm 0.09$ &  $<0.21$ & $<0.53$ \\ 
I(N)-SMA4 & 17 20 54.69 & -35 45 08.5 & $0.17$ & $0.33 \pm 0.07$ &  $<0.70$ & $<0.77$ \\ 
I(N)-SMA5 & 17 20 55.08 & -35 45 02.0 & $0.27$ & $0.28 \pm 0.07$ &  $<0.12$ & $<0.63$ \\ 
I(N)-SMA6 & 17 20 54.59 & -35 45 17.9 & $0.27$ & $0.47 \pm 0.10$ &  $<0.29$ & $<0.69$ \\ 
I(N)-SMA7 & 17 20 54.96 & -35 44 57.3 & $0.07$ & $0.20 \pm 0.05$ &  $<0.51$ & $<0.68$ \\ 
\hline 
I(N) Total  &         &              &  & $4.6 \pm 0.9$\tablenotemark{b} & \\ 
\hline
\end{tabular}
\centerline{$^{\rm a}$ The extended nebula of the compact HII region precludes a meaningful infrared point 
source upper limit}
\centerline{$^{\rm b}$ The total 1.3mm flux density includes the extended emission; the listed uncertainties 
include 20\% calibration uncertainty}

\end{table*}

\subsection{\ngcin} 

\begin{figure}
\includegraphics[scale=0.42,angle=0]{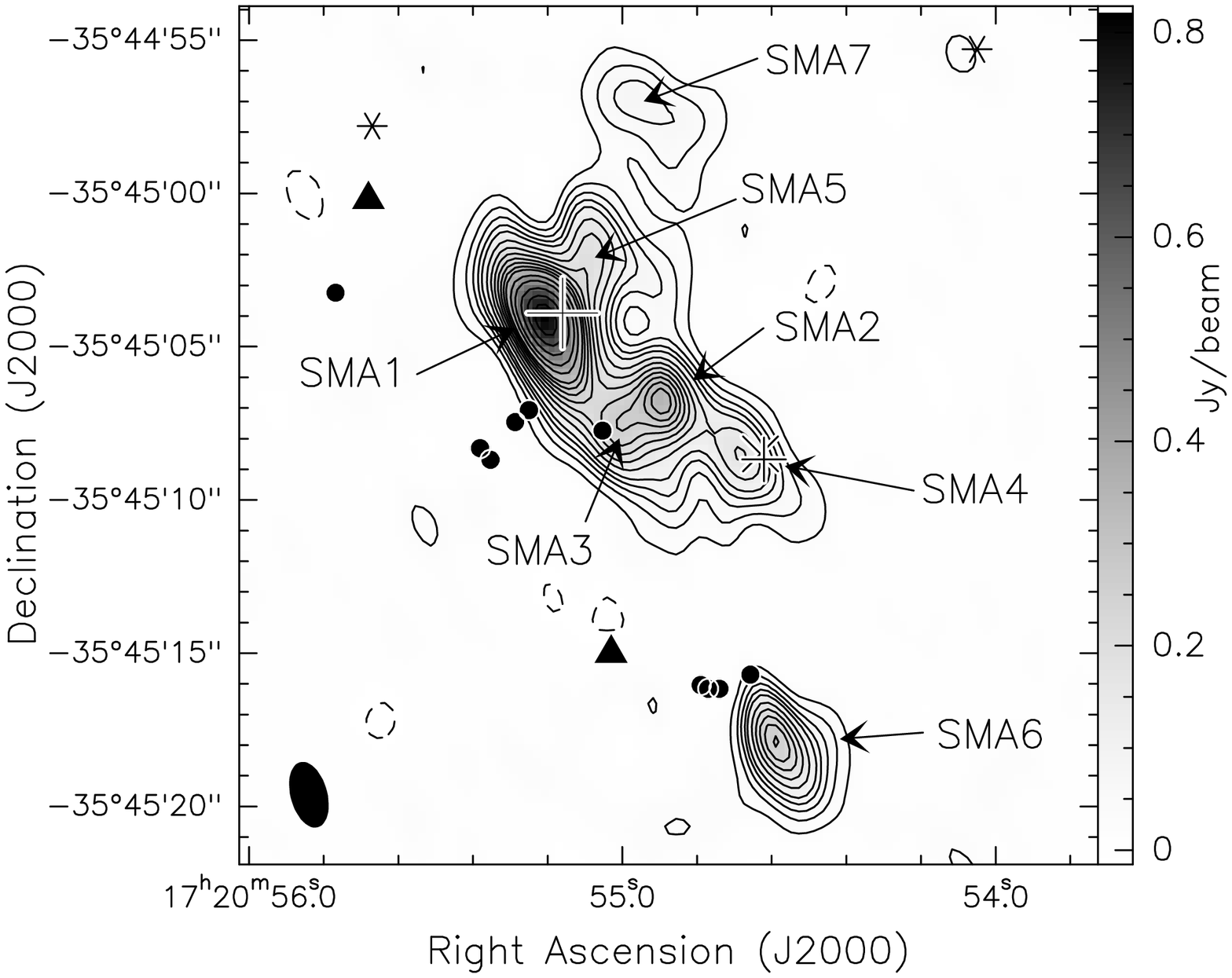}
\caption{
1.3~mm continuum image of \ngcin. 
Contour levels are (-3, 3, 6, 9, 12, 15, 20, 25, 30, 35, 40, 45, 50, 60, 70,  
85, 100, 115) $\times$ 6.5 mJy/beam.  Filled circles mark 44~GHz
Class I \methanol\ masers \citep{Kogan98}, the small cross marks 
a 6.7~GHz Class II \methanol\ maser \citep{Norris93,Walsh98}, the X 
marks the faint 3.6~cm source \citep{Carral02}, and 
the 6-pointed stars mark infrared H$_2$ knots \citep{Megeath99}.  
The filled triangles mark 25~GHz Class I \methanol\ masers and the 
large cross indicates the \ammonia\ peak \citep{Beuther05}.
\label{fig2}}
\end{figure}

\begin{figure}
\includegraphics[scale=0.5,angle=0]{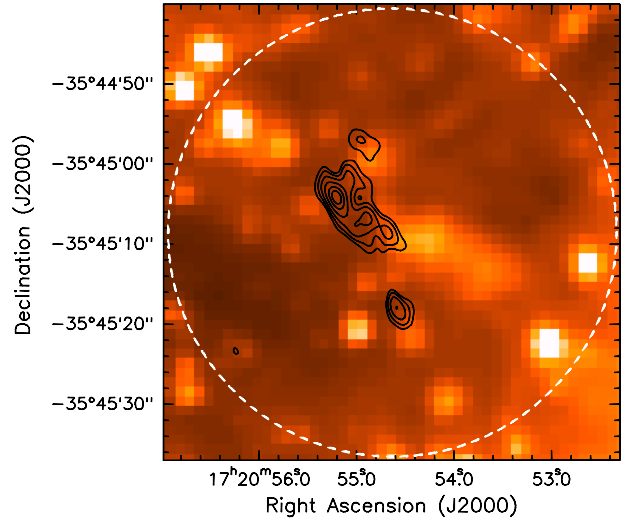}
\caption{$4.5\mu$m IRAC image of \ngcin. Contours indicate the
1.3mm continuum emission (5, 10, 20, 40, 70, 100 $\times$ 6.5~mJy/beam)
and the dashed circle shows the primary beam of the 1.3mm SMA observations.
The newly-detected infrared nebula happens to lie near the center of 
this circle, just west of I(N)-SMA4.  
\label{fig3}}
\end{figure}

We have resolved the millimeter continuum of \ngcin\ into seven
sources, which we denote as I(N)-SMA1..7 (Figure~\ref{fig2}).
Including the extended emission, the fraction of single-dish flux
recovered in our image is $32 \pm 8$\%.  Nearly half of the total
emission originates from I(N)-SMA1.  Although this field lacks any
strong centimeter continuum emission, two faint 3.6 cm (0.3~mJy)
sources have been reported \citep{Carral02}.  One of these sources
lies within $0.6''$ of I(N)-SMA4, which is also the source closest to
a Class II \methanol\ maser \citep{Walsh98,Norris93}.  The faint
continuum source with its (probable) associated class II \methanol\
maser may represent a hypercompact HII region, similar to the sources
found by \citet{vandertak05} in other massive star formation regions.
If this interpretation is true, then this object is an important power
source in the I(N) region.  Class I \methanol\ masers have also been
identified in this region at 25~GHz \citep{Beuther05} and at 44~GHz
but with uncertain astrometry \citep{Kogan98}.  None of the
\millimeter\ sources have counterparts in our IRAC data ($5\sigma$
upper limits at $4.5\mu$m and $8\mu$m are given in
Table~\ref{table1}).  However, directly west of I(N)-SMA4 is an
extended infrared source detected at 4.5 and 5.8~$\mu$m
(Figure~\ref{fig3}).  The elongated morphology suggests a jet or a
reddened reflection nebulosity from an outflow cavity.  In either
scenario, the extended mid-IR emission would be the result of outflow
activity from the embedded source in I(N)-SMA4. The unipolarity could
be explained if the other lobe of the putative outflow was obscured by
stronger extinction in its direction.  This possible association is
the only evidence, albeit indirect, for the I(N)-SMA sources at
wavelengths $\le 8$~$\mu$m.

\subsection{Mass estimates from the dust emission} 

We can estimate the mass of the individual dust sources by placing our
millimeter continuum measurements in the context of a simple
isothermal model of optically-thin dust emission \citep{Beltran06}.
The assumption of low optical depth is justified because the peak
continuum brightness temperature of the strongest dust source (I-SMA1)
is only 6.3~K.  The input parameters to the model include the dust
temperature, the dust mass opacity coefficient at the observed
wavelength ($\kappa_{\rm 1.3mm}$), and the gas-to-dust mass ratio.
Based on our SMA spectra (Hunter et al., in preparation), we can
categorize the sources into three groups: A) strong hot-core line
emission: I-SMA1, I-SMA2 and I(N)-SMA1, B) weak line emission: I-SMA3
and I(N)-SMA2, and C) nearly line-free emission.  Thus, from a
qualitative viewpoint, the gas and dust is likely to be warmest in the
group A sources, and coolest in the group C sources.  By combining
this information with the various gas temperature measurements from
the literature, we have assigned a probable value for the dust
temperature of each source including upper and lower limits (see
Table~\ref{table2}).  

\begin{table}
\centering
\caption{Range of estimated temperatures and masses of dust cores in \ngci\ and I(N) \label{table2}}  
\begin{tabular}{cccccccc}  
\hline\hline
       & & \multicolumn{2}{c}{cold limit} & \multicolumn{2}{c}{nominal} & \multicolumn{2}{c}{warm limit} \\ 
       & $\kappa$          & $T$  & $M$      & $T$ & $M$         & $T$ & $M$\\
Source & (cm$^2$~g$^{-1}$) & (K)  & ($M_\odot$) & (K) & ($M_\odot$) & (K) & ($M_\odot$)\\
\hline
I-SMA1 & 2 & 75 &  23 & 100 & 17 & 300 &  5.4 \\
I-SMA2 & 2 & 75 &  15 & 100 & 11 & 300 &  3.5 \\
I-SMA3 & 1 & 40 &  52 &  60 & 33 & 100 & 19   \\
I-SMA4 & 1 & 20 &  66 &  33 & 36 &  50 & 22   \\
\hline  %
Total  &   &    & 156 &     & 97 &     & 50   \\
\hline
I(N)-SMA1 & 2 &  65 & 15  & 100 & 10  & 300 & 3.1 \\
I(N)-SMA2 & 1 &  20 & 30  &  40 & 13  & 100 & 4.8 \\
I(N)-SMA3 & 1 &  20 & 23  &  33 & 13  &  50 & 7.8 \\
I(N)-SMA4 & 1 &  20 & 20  &  33 & 11  &  50 & 6.1 \\
I(N)-SMA5 & 1 &  20 & 17  &  33 &  9  &  50 & 5.2 \\
I(N)-SMA6 & 1 &  20 & 28  &  33 & 15  &  50 & 8.7 \\
I(N)-SMA7 & 1 &  20 & 12  &  33 &  6  &  50 & 3.7 \\
\hline %
I(N) Total     &   &     & 145 &     & 77  &     & 39 \\
\hline
\end{tabular}
\end{table}

For the sources in group A, the brightness
temperature of the arcsecond scale \ammonia\ (2,2) emission provides
the lower-limit temperature \citep{Beuther05}.  For I-SMA1 and I-SMA2,
we set the ``nominal'' temperature equal to the dust temperature
(100~K) obtained from fits to the single-dish spectral energy
distribution (SED) for source I \citep{Sandell00}.  We also use 100~K
for I(N)-SMA1, because the single-dish intensity ratio of \ammonia\
(6,6) to (3,3) indicates the presence of gas with $T>95$~K
\citep{Kuiper95}; furthermore, recent interferometric measurements
have localized the bulk of the \ammonia\ (6,6) emission to the
position of I(N)-SMA1 (Beuther et al., in preparation). For an upper
limit temperature for group A, we adopt 300~K because an excitation
temperature of 295~K was measured for H$_2$CO in \ngci\ by
\citet{Mangum93} and 213~K for \methanol\ by \citet{vandertak03}.  For
the sources in group B, we set the nominal temperature equal to the CO
excitation temperature derived by \citet{Kraemer99} which is 60~K for
\ngci\ and 40~K for I(N).  For the sources in group C, we set the
nominal temperature equal to the value (33~K) derived from a Large
Velocity Gradient analysis of a single-dish submillimeter line survey
\citep{McCutcheon00}. In both I and I(N), the lower limit temperature
of the group B and C sources is taken to be the temperature of the
coolest dust core (20~K) in the surrounding region as measured by
\citet{Sandell00}, while the upper limit is taken to be 50~K due to
the lack of direct evidence of any warmer gas at these positions.
Measurements of $\kappa_{\rm 1.3mm}$ in \ngci\ and I(N) with a $30''$
beam yield values from 1.0 to $1.2 \pm 0.6$~cm$^2$~g$^{-1}$
\citep{Schwartz89}.  However, due to the large range of temperature in
the compact millimeter sources, we have chosen to use tabulated values
of $\kappa_{\rm 1.3mm}$ from \citet{Ossenkopf94} for a density of
$10^6$~cm$^{-3}$.  For sources in group A, we use a $\kappa_{\rm
1.3mm} = 2$~cm$^2$~g$^{-1}$ for dust grains without mantles, which is
appropriate for regions where protostellar heating has destroyed the
solid ice but has not yet dispersed the dust aggregates.  In the other
millimeter sources which show no hot core emission and are opaque at
infrared wavelengths, the grains are likely to have thick ice mantles;
thus we use $\kappa_{\rm 1.3mm} = 1$~cm$^2$~g$^{-1}$.  Applying a
gas-to-dust mass ratio ($g$) of 100 \citep{Sodroski97}, these
assumptions yield the masses listed in Table~\ref{table2}.  Due to
uncertainties in $\kappa_{\rm 1.3mm}$ and $g$, the uncertainties in
all of the masses in Table~\ref{table2} are likely to be at least a
factor of two.  Subject to these uncertainties, the total mass
estimates of the compact objects (in both the warm and cold
temperature limits) are of similar magnitude in I and I(N), ranging
from $\sim 50$~\msun\ to $\sim 150$~\msun.  In I(N), this mass is
divided among a larger number sources which are, in turn, spread over
a larger angular extent.

\section{Discussion} 

The observed multiplicity and strength of the \millimeter\ continuum
emission is quite similar between \ngciabbrev\ and I(N), in contrast
to their strikingly different appearance in the NIR \citep{Megeath99}
and in thermal lines of \ammonia\ and \methanol\
\citep{Beuther05}. Although previous maser and \ammonia\ observations
of \ngci\ had suggested the presence of multiple sources, our
\millimeter\ image provides a direct and unambiguous picture of a
massive protocluster.  The estimated mass of interstellar material
toward each source is sufficient to form a massive star and is
significantly greater than typical objects detected in the single-dish
surveys of low-mass star-forming regions such as Perseus, where an
average core mass of 2.3~\msun\ has been found with a beamsize of
$\sim 8000$~AU \citep{Enoch06}.  In addition, it is likely that a
compact protostellar object may already be present at the center of
each source, heating the surrounding gas and dust and leading to our
strong millimeter detections.  This scenario is particularly likely
for the strong hot-core sources I-SMA1 and I-SMA2.  It is important to
note that the masses in Table~\ref{table2} do not include the total
stellar mass that may be present in addition to the interstellar
material.

In the case of \ngcin, the presence of outflows, masers, warm gas
emission, and a faint centimeter continuum source also indicates the
presence of embedded sources.  In comparison to source I, this cluster
of sources may be in an earlier evolutionary phase, as suggested by
the lack of a NIR cluster, HII region, or bright mid-IR source, and
the significantly higher gas column density inferred from \ntwohplus\
observations \citep{Pirogov03}. However, if the millimeter sources
contain central protostars, then the lack of mid-IR detections toward
all seven sources is especially curious, because, in contrast to
source I, there is no confusion from bright, extended nebulosity in
this region.  If we consider the I(N) bolometric luminosity of
$1.9~\times~10^3$~\lsun \citep{Sandell00}, the earliest ZAMS star that
could be present is type B2V \citep{Hanson97}.  Our lowest $4.5\mu$m
upper limit of 0.11~mJy (from Table~\ref{table1}) corresponds to an
apparent magnitude $m_{\rm {4.5\mu}m} > 15.5$.  This limit is
consistent with the non-detection of an embedded B2V ZAMS star as long
as the extinction at $4.5\mu$m is $\geq 4.9$~mag, assuming an absolute
stellar magnitude of $M_{K} = -0.68$ \citep{Hanson97} and an intrinsic
color $(M_{K} - M_{\rm 4.5{\mu}m}) = -0.08$~mag \citep{Koornneef83}.
This level of extinction at $4.5\mu$m corresponds to 9.3~mag of
$K$~band extinction \citep{Indebetouw05}, 83~mag of $V$~band
extinction \citep{Rieke85}, and a total hydrogen column density of
$1.8 \times 10^{23}~{\rm cm}^{-2}$ \citep{Ryter96}.  Even using the
smallest mass estimates from Table~\ref{table2}, the gas column
density implied by the dust emission is several times larger than this
value and can easily explain the non-detections.  As a result, it is
unclear whether these \millimeter\ sources contain young protostars
(B2 or later) that are simply heavily obscured, or whether they are
intrinsically faint in the mid-IR due to their youth, and may
eventually evolve into higher luminosity O stars.  In either case, the
high angular resolution and sensitivity of our observations have
revealed the presence of compact systems where intermediate to high
mass stars are being formed.

In \ngci, all four of the compact millimeter continuum sources are
concentrated within a region of projected radius $\leq 0.05$~pc which
is significantly smaller than the 0.33~pc radius of the NIR cluster
\citep{Tapia96} and the 0.23~pc radius of the search region (i.e. the
SMA primary beam).  Furthermore, the millimeter sources in \ngci\
reside close to the geometric center of the NIR cluster where the
number density of millimeter sources (8000~pc$^{-3}$) exceeds the
stellar density of the NIR cluster (1200~pc$^{-3}$).  These findings
suggest that massive star formation is biased toward the center of the
parent cluster.  The projected separation of the four components of
\ngci\ range from 5800 to 11200~AU. Because the ratio of distances
between all components is less than three, it satisfies the criterion
for identifying Trapezia as described by \citet{Abt00} and originated
by \citet{Ambartsumian54}.  If the total luminosity of \ngci\
($2.6\times 10^5$ \lsun) was apportioned equally to four sources,
their individual luminosities would correspond to that of an O8.5 ZAMS
star \citep{Hanson97}.  These findings argue that the \ngci\ SMA
sources may constitute a proto-Trapezium system, similar to the more
evolved Orion Nebula Trapezium (10,000~AU in diameter) and many other
optically visible nebulae and clusters.  Although the I(N) SMA sources
are more spread out, five of the sources of I(N) reside inside a
region of 10,000~AU, suggesting that a similar system is forming in
I(N).  These observations suggest that the presence of Trapezia in
clusters and nebulae may result from the massive stars forming
preferentially in the centers of clusters in the deepest part of the
gravitational well.  

Are these new observations consistent with previous claims
\citep{Megeath99,Sollins04} that \ngcin\ is in an earlier stage of
star formation?  In source I, we have identified examples of young
massive protostars in the center of an established infrared cluster
containing low to intermediate mass stars.  In source I(N), we see
evidence of massive protostars without a surrounding NIR cluster,
which challenges the common picture of high mass stars forming after
the first generation of low mass stars (e.g. \citet{Herbig62};
\citet{Kumar06}).  More sensitive infrared imaging is needed to probe
deeper through the extinction for signs of lower mass protostars in
this field.  However, until such evidence is found, the question of
the relative age of I and I(N) hinges on three main facts: 1) the
ratio of bolometric luminosity between I and I(N) is large ($\approx
140$) \citep{Sandell00}; 2) the I(N) region exhibits significantly
less hot-core line emission than \ngci\ (Thorwirth et al. (2003);
Hunter et al., in preparation); 3) a cluster of compact dust continuum
sources exists in I and I(N) with a comparable amount of mass (this
paper).  These observations suggest that \ngci\ is a more evolved
cluster than I(N), or that I(N) is forming a cluster with a larger
number of stars but of lower mass.  In either case, our millimeter
data provide strong evidence that \ngcin\ is forming a cluster of
stars, even though an associated NIR cluster has not been identified,
and may be still in the process of forming.  Finally, we note that the
virial masses derived from single-dish molecular line spectra
(including \ntwohplus\ by \citet{Pirogov03} and HC$_3$N by
\citet{Sollins04}) for both \ngciabbrev\ and I(N) are several times
higher than the total estimated mass contained in the compact
\millimeter\ sources.  This fact illustrates the wealth of
star-forming material in both regions that has either not assembled
into compact protostars or exists in a wider distribution of smaller
cores below our sensitivity limit.  It is this gas which may be
forming the cluster surrounding the SMA sources in \ngcin\, and which
may further increase the population of the cluster in \ngci.

\bigskip

\acknowledgments

The authors wish to thank the anonymous referee for a diligent report
which improved this manuscript.  This work is based in part on
observations made with the {\it Spitzer} Space Telescope, which is
operated by the Jet Propulsion Laboratory, California Institute of
Technology under a contract with NASA.  This research has made use of
NASA's Astrophysics Data System Bibliographic Services and the SIMBAD
database operated at CDS, Strasbourg, France.  Support for for STM was
provided by NASA through contract 1256790 issued by JPL/Caltech. HB
acknowledges financial support by the Emmy-Noether-Program of the
Deutsche Forschungsgemeinschaft (DFG, grant BE2578).


\begin{thebibliography}{37}
\expandafter\ifx\csname natexlab\endcsname\relax\def\natexlab#1{#1}\fi

\bibitem[{{Abt} \& {Corbally}(2000)}]{Abt00} 
{Abt}, H.~A. \& {Corbally}, C.~J. 2000, \apj, 541, 841

\bibitem[Ambartsumian(1954)]{Ambartsumian54} Ambartsumian, V.~A.\ 
1954, Les Processus Nucl{\'e}aires dans les Astres, 293 
 
\bibitem[Beltr{\'a}n et al.(2006)]{Beltran06} Beltr{\'a}n, M.~T., 
Brand, J., Cesaroni, R., Fontani, F., Pezzuto, S., Testi, L., \& Molinari, 
S.\ 2006, \aap, 447, 221 

\bibitem[{{Beuther} {et~al.}(2005){Beuther}, {Thorwirth}, {Zhang}, {Hunter},
  {Megeath}, {Walsh}, \& {Menten}}]{Beuther05}
{Beuther}, H., {Thorwirth}, S., {Zhang}, Q., {Hunter}, T.~R., {Megeath}, S.~T.,
  {Walsh}, A.~J., \& {Menten}, K.~M. 2005, \apj, 627, 834

\bibitem[{{Blundell}(2004)}]{Blundell04}
{Blundell}, R. 2004, in {Proceedings of the 15th International Symposium on
  Space Terahertz Technology}, 3

\bibitem[Bonnell \& Davies(1998)]{Bonnell98} 
Bonnell, I.~A., \&  Davies, M.~B.\ 1998, \mnras, 295, 691 

\bibitem[{{Carral} {et~al.}(1997){Carral}, {Kurtz}, {Rodr{\'{\i}}guez}, {de Pree}, \&
  {Hofner}}]{Carral97}
{Carral}, P., {Kurtz}, S.~E., {Rodr{\'{\i}}guez}, L.~F., {de Pree}, C., \& {Hofner},
  P. 1997, \apjl, 486, L103+

\bibitem[{{Carral} {et~al.}(2002){Carral}, {Kurtz}, {Rodr{\'{\i}}guez},
  {Menten}, {Cant{\'o}}, \& {Arceo}}]{Carral02}
{Carral}, P., {Kurtz}, S.~E., {Rodr{\'{\i}}guez}, L.~F., {Menten}, K.,
  {Cant{\'o}}, J., \& {Arceo}, R. 2002, \aj, 123, 2574

\bibitem[{{Cheung} {et~al.}(1978){Cheung}, {Frogel}, {Hauser}, \& {Gezari}}]{Cheung78}
{Cheung}, L., {Frogel}, J.~A., {Hauser}, M.~G., \& {Gezari}, D.~Y., \apjl, 
  226, L149

\bibitem[{{De Buizer} {et~al.}(2000){De Buizer}, {Pi{\~n}a}, \&
  {Telesco}}]{DeBuizer00}
{De Buizer}, J.~M., {Pi{\~n}a}, R.~K., \& {Telesco}, C.~M. 2000, \apjs, 130,
  437

\bibitem[{{De Buizer} {et~al.}(2002){De Buizer}, {Radomski}, {Pi{\~n}a}, \&
  {Telesco}}]{DeBuizer02}
{De Buizer}, J.~M., {Radomski}, J.~T., {Pi{\~n}a}, R.~K., \& {Telesco}, C.~M.
  2002, \apj, 580, 305

\bibitem[Emerson et al.(1973)]{Emerson73} 
Emerson, J.~P., Jennings, R.~E., \& Moorwood, A.~F.~M.\ 1973, \apj, 184, 401 

\bibitem[Enoch et al.(2006)]{Enoch06} 
Enoch, M.~L., et al. 2006, \apj, 638, 293 

\bibitem[{{Fazio} {et~al.}(2004)}]{Fazio04}
{Fazio}, G.~G. {et~al.} 2004, \apjs, 154, 10

\bibitem[{{Forster}(1992)}]{Forster92}
{Forster}, J. 1992, in {Astrophysical Masers}, ed. A.~{Clegg} \& G.~{Nedoluha}
  (Springer-Verlag)

\bibitem[{{Gezari}(1982)}]{Gezari82}
{Gezari}, D.~Y. 1982, \apjl, 259, L29

\bibitem[Hanson et al.(1997)]{Hanson97} 
Hanson, M.~M., Howarth, I.~D., \& Conti, P.~S.\ 1997, \apj, 489, 698 

\bibitem[Herbig(1962)]{Herbig62} 
Herbig, G.~H.\ 1962, Advances in Astronomy and Astrophysics, 1, 47 

\bibitem[Indebetouw et al.(2005)]{Indebetouw05} 
Indebetouw, R., et al.\ 2005, \apj, 619, 931 

\bibitem[{{Kogan} \& {Slysh}(1998)}]{Kogan98}
{Kogan}, L. \& {Slysh}, V. 1998, \apj, 497, 800

\bibitem[Koornneef(1983)]{Koornneef83} 
Koornneef, J.\ 1983, \aap, 128, 84 

\bibitem[{{Kraemer} \& {Jackson}(1995)}]{Kraemer95}
{Kraemer}, K.~E. \& {Jackson}, J.~M. 1995, \apjl, 439, L9

\bibitem[Kraemer et al.(1999)]{Kraemer99} Kraemer, K.~E., 
Deutsch, L.~K., Jackson, J.~M., Hora, J.~L., Fazio, G.~G., Hoffmann, W.~F., 
\& Dayal, A.\ 1999, \apj, 516, 817 

\bibitem[{{Kuiper} {et~al.}(1995){Kuiper}, {Peters}, {Forster}, {Gardner}, \&
  {Whiteoak}}]{Kuiper95}
{Kuiper}, T.~B.~H., {Peters}, W.~L., {Forster}, J.~R., {Gardner}, F.~F., \&
  {Whiteoak}, J.~B. 1995, \apj, 446, 692

\bibitem[Kumar et al.(2006)]{Kumar06} 
Kumar, M.~S.~N., Keto, E., \& Clerkin, E.\ 2006, \aap, 449, 1033 

\bibitem[{{Mangum} \& {Wootten}(1993)}]{Mangum93}
{Mangum}, J.~G. \& {Wootten}, A. 1993, \apjs, 89, 123

\bibitem[{{McBreen} {et~al.}(1979)}]{McBreen79}
{McBreen}, B., {Fazio}, G.~G., {Stier}, M., \& {Wright}, E.~L. 1979, \apjl, 
  232, L138

\bibitem[{{McCutcheon} {et~al.}(2000){McCutcheon}, {Sandell}, {Matthews},
  {Kuiper}, {Sutton}, {Danchi}, \& {Sato}}]{McCutcheon00}
{McCutcheon}, W.~H., {Sandell}, G., {Matthews}, H.~E., {Kuiper}, T.~B.~H.,
  {Sutton}, E.~C., {Danchi}, W.~C., \& {Sato}, T. 2000, \mnras, 316, 152

\bibitem[{{Megeath} \& {Tieftrunk}(1999)}]{Megeath99}
{Megeath}, S.~T. \& {Tieftrunk}, A.~R. 1999, \apjl, 526, L113

\bibitem[{{Megeath} {et~al.}(2005){Megeath}, {Wilson}, \& {Corbin}}]{Megeath05}
{Megeath}, S.~T., {Wilson}, T.~L., \& {Corbin}, M.~R. 2005, \apjl, 622, L141

\bibitem[{{Migenes} {et~al.}(1999){Migenes}, {Horiuchi}, {Slysh}, {Val'tts},
  {Golubev}, {Edwards}, {Fomalont}, {Okayasu}, {Diamond}, {Umemoto}, {Shibata},
  \& {Inoue}}]{Migenes99}
{Migenes}, V., {Horiuchi}, S., {Slysh}, V.~I., {Val'tts}, I.~E., {Golubev},
  V.~V., {Edwards}, P.~G., {Fomalont}, E.~B., {Okayasu}, R., {Diamond}, P.~J.,
  {Umemoto}, T., {Shibata}, K.~M., \& {Inoue}, M. 1999, \apjs, 123, 487

\bibitem[{{Neckel}(1978)}]{Neckel78}
{Neckel}, T. 1978, \aap, 69, 51

\bibitem[{{Norris} {et~al.}(1993){Norris}, {Whiteoak}, {Caswell}, {Wieringa},
  \& {Gough}}]{Norris93}
{Norris}, R.~P., {Whiteoak}, J.~B., {Caswell}, J.~L., {Wieringa}, M.~H., \&
  {Gough}, R.~G. 1993, \apj, 412, 222

\bibitem[{{Ossenkopf} \& {Henning}(1994)}]{Ossenkopf94}
{Ossenkopf}, V. \& {Henning}, T. 1994, \aap, 291, 943

\bibitem[{{Persi} {et~al.}(1996){Persi}, {Roth}, {Tapia}, {Marenzi}, {Felli},
  {Testi}, \& {Ferrari-Toniolo}}]{Persi96}
{Persi}, P., {Roth}, M., {Tapia}, M., {Marenzi}, A.~R., {Felli}, M., {Testi},
  L., \& {Ferrari-Toniolo}, M. 1996, \aap, 307, 591

\bibitem[{{Persi} {et~al.}(1998){Persi}, {Tapia}, {Felli}, {Lagage}, \&
  {Ferrari-Toniolo}}]{Persi98}
{Persi}, P., {Tapia}, M., {Felli}, M., {Lagage}, P.~O., \& {Ferrari-Toniolo},
  M. 1998, \aap, 336, 1024

\bibitem[{{Pirogov} {et~al.}(2003){Pirogov}, {Zinchenko}, {Caselli},
  {Johansson}, \& {Myers}}]{Pirogov03}
{Pirogov}, L., {Zinchenko}, I., {Caselli}, P., {Johansson}, L.~E.~B., \&
  {Myers}, P.~C. 2003, \aap, 405, 639

\bibitem[Rieke \& Lebofsky(1985)]{Rieke85} Rieke, G.~H., \& 
Lebofsky, M.~J.\ 1985, \apj, 288, 618 

\bibitem[Rodr{\'{\i}}guez et al.(1982)]{Rodriguez82} 
Rodr{\'{\i}}guez, L.~F., Canto, J., \& Moran, J.~M.\ 1982, \apj, 255, 103

\bibitem[Ryter(1996)]{Ryter96} 
Ryter, C.~E.\ 1996, \apss, 236, 285 
 
\bibitem[{{Sandell}(2000)}]{Sandell00}
{Sandell}, G. 2000, \aap, 358, 242

\bibitem[{{Schwartz} {et~al.}(1989){Schwartz}, {Snell}, \&
  {Schloerb}}]{Schwartz89}
{Schwartz}, P.~R., {Snell}, R.~L., \& {Schloerb}, F.~P. 1989, \apj, 336, 519

\bibitem[Sharpless(1954)]{Sharpless54} 
Sharpless, S.\ 1954, \apj, 119, 334 

\bibitem[Sodroski et al.(1997)]{Sodroski97} 
Sodroski, T.~J., Odegard, N., Arendt, R.~G., Dwek, E., Weiland, J.~L., H
auser, M.~G., \& Kelsall, T.\ 1997, \apj, 480, 173 

\bibitem[Sollins \& Megeath(2004)]{Sollins04} 
Sollins, P.~K., \& Megeath, S.~T.\ 2004, \aj, 128, 2374 

\bibitem[{{Straw} \& {Hyland}(1989)}]{Straw89a}
{Straw}, S.~M. \& {Hyland}, A.~R. 1989, \apj, 340, 318

\bibitem[{{Tapia} {et~al.}(1996){Tapia}, {Persi}, \& {Roth}}]{Tapia96}
{Tapia}, M., {Persi}, P., \& {Roth}, M. 1996, \aap, 316, 102

\bibitem[{{Thorwirth} {et~al.}(2003){Thorwirth}, {Winnewisser}, {Megeath}, \&
  {Tieftrunk}}]{Thorwirth03}
{Thorwirth}, S., {Winnewisser}, G., {Megeath}, S.~T., \& {Tieftrunk}, A.~R.
  2003, in ASP Conf. Ser. 287: Galactic Star Formation Across the Stellar Mass
  Spectrum, 257--260

\bibitem[{{van der Tak} {et~al.}(2003){van der Tak}, {Boonman}, {Braakman}, \&
  {van Dishoeck}}]{vandertak03}
{van der Tak}, F.~F.~S., {Boonman}, A.~M.~S., {Braakman}, R., \& {van
  Dishoeck}, E.~F. 2003, \aap, 412, 133

\bibitem[{van der Tak} \& Menten(2005)]{vandertak05} 
van der Tak, F.~F.~S., \& Menten, K.~M.\ 2005, \aap, 437, 947 

\bibitem[{{Walsh} {et~al.}(1998)}]{Walsh98}
{Walsh}, A.J., {Burton}, M.~G., {Hyland}, A.~R., \& {Robinson}, G. 1998, 
   \mnras, 301, 640

\end{thebibliography}
\end{document}